\documentstyle[11pt,newpasp,twoside,epsf]{article}
\def\edcomment#1{\iffalse\marginpar{\raggedright\sl#1\/}\else\relax\fi}
\reversemarginpar

\begin{document}
\title{M\,31's molecular arms at all scales to below 10\,pc}

\author{N. Neininger}
\affil{Radioastronomisches Institut der Universit\"at Bonn, 
Auf dem H\"ugel 71, D-53121 Bonn, Germany}  

\author{in collaboration with:\quad Ch.\ Nieten$^1$,$\;$
M. Gu\'elin$^2$,$\;$  H. Ungerechts$^3$,$\;$ R. Lucas$^2$,$\;$
S. Muller$^2$,$\;$ R. Wielebinski$^1$} 
\affil{$^1$MPIfR, D-53121 Bonn (Germany), $^2$IRAM, F-38406
St.\,Martin d'H\`eres (France), $^3$IRAM E-18012 Granada (Spain)}

\begin{abstract}
I present a unique data set for the study of molecular gas in
galaxies: a complete, high-resolution survey of the CO in M\,31 and
additional local studies. The fully sampled survey has an angular
resolution of $23''$\,FWHM and the interferometric data attain the
pc-scale with sub-arcsecond resolution. For the first time it is now
possible to study large and small scales in conjunction. Thus we are
able to derive the global structure and study the links down to the
individual cloud complexes and star formation regions.
\end{abstract}

\section{Overview}

We are pursuing a project to study the relationship between molecular
clouds and star formation in a single object and for the widest
possible range of scales. This needs high sensitivity and complete
sampling at high angular resolutions. The best spatial resolution
can of course be obtained in our Milky Way, but it is very difficult
to investigate the interplay between the various constituents of the
interstellar medium (ISM) here. This is due to our ``inside view''
that makes distance determinations difficult and largely impedes the
construction of a global picture of the structure of our Galaxy. The
obvious choice for the object thus is a nearby external galaxy and we
have started to investigate the molecular content of the closest large
spiral: M\,31.
 
Of course, there have been numerous earlier attempts to derive the
distribution of the molecular gas in M\,31, but this turned out to be
difficult and the first complete survey was published only a few years
ago (Dame et al.\ 1993). Observational techniques new in the field of
millimeter-astronomy, such as the On-The-Fly (OTF) method, now allow
to map such large areas fully sampled in a reasonably short time. Our
new map (cf.\ Fig.~1) clearly shows that the CO emission of M\,31
covers only a small percentage of the disk and thus both, a complete
coverage and a well matched angular resolution are necessary.

Once the distribution of the giant molecular cloud complexes (GMCs)
was known, we started detailed observations at the highest possible
angular resolutions. Several regions were chosen and investigated with
the sensitive Plateau de Bure interferometer (PdBI). Of particular
interest are complexes with peculiarities in their spectra. In
contrast to the Galaxy, the kinematical information of the spectra in
M\,31 can be linked directly to a position. The observed velocity
anomalies hence indeed indicate local disturbances.

\section{Observations}
\subsection{The large-scale survey}

The first step for a thorough investigation of the molecular gas
content is a complete survey of the entire disk at sufficiently high
angular resolution. We used an OTF mapping procedure implemented at
the IRAM 30-m telescope to cover an area of about 1 square degree in
the $^{12}$CO(1-0) transition with a beam of 23\arcsec{} FWHM. The
whole area is covered at least twice, and sampled on a 4\arcsec{}
grid. Several smaller regions of interest have been mapped on a
2\arcsec{} grid (and correspondingly longer integration times per
beam) to obtain high quality data of the (2-1) transition as well.
Details about the mapping procedure and the data reduction are 
given in Neininger (2000).

\begin{figure}[ht]
\plotfiddle{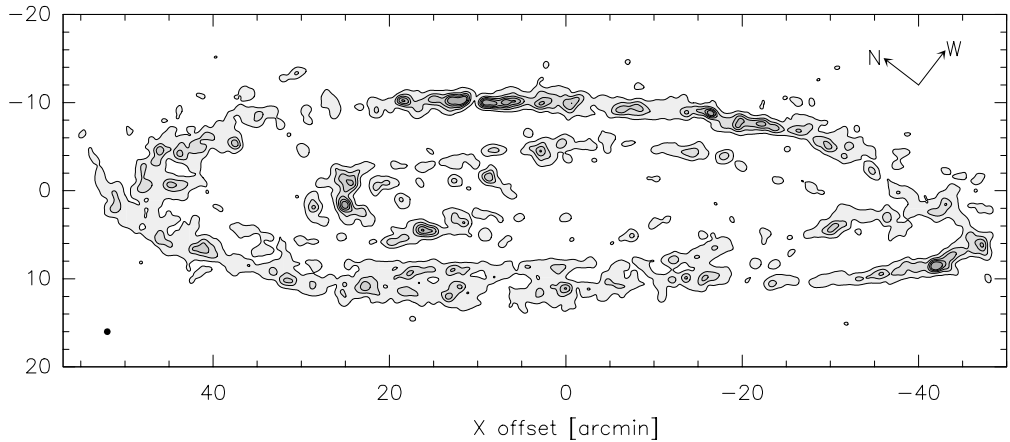}{54mm}{90}{56}{56}{257}{-24}
\caption{The emission of the $^{12}$CO(1-0) in M\,31 at an angular
resolution of 45\arcsec{} FWHM (indicated by the dot in the lower left
corner). The lowest contour is at a level of at least $4\sigma$
RMS. The molecular gas is concentrated in thin arms. Nevertheless, a
few isolated clouds can be found in the ``interarm'' regions. Emission
has been found between radii of about 4\,kpc and 18\,kpc. 
} 
\end{figure}

\noindent
The map (Fig.~1) shows the CO emission concentrated in narrow
arm segments that tend to break up into filamentary pieces. The
arm/interarm contrast is about $10:1$ -- much higher than that of the
atomic gas. No emission is found inside of about 4\,kpc radius at the
noise level of the survey; close to the nucleus, one cloud complex has
been detected at an about 5 times lower intensity, using a special
detection strategy (Melchior et al.\ 2000).

\subsection{Interferometric studies}

For the PdBI observations, we choose GMCs or GMC groups that spread
over the area of a few PdBI primary beams while representing a broad
variety of the cloud complexes found. The innermost cloud in this
sample is located at a radius of 5\,kpc, the outermost group at
$\sim18$\,kpc. In addition, we choose other GMCs with various
morphologies in different environments, including some with multiple
spectral components. Up to now, our sample consists of six regions
covered with mosaicked observations of 2 to 9 pointings -- mainly in
the south-western part of the galaxy, but also containing the prominent
association of molecular gas on the northern major axis at a distance
of 5\,kpc from the center.

\begin{figure}
\vbox{
\plotfiddle{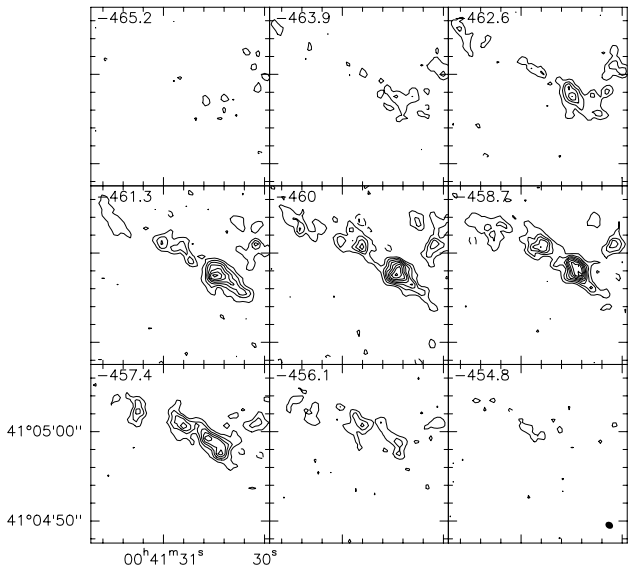}{49mm}{270}{40}{40}{-200}{150} \vspace{-51mm}}
\hfill\parbox[b]{60mm}{
\caption
{Channel maps of the $^{12}$CO(2-1) emission of a GMC at 5\,kpc
radius. The beam size is $0.9''\times0.74''$ ($\simeq3.5\,$pc),
indicated in the lower right corner. Each box is marked with its
velocity channel, coordinates are J2000.0. No hint at the obvious
substructure is found in the corresponding survey data. Even for this
quiescent cloud the virial masses determined from the two data sets
differ significantly.}  }
\end{figure}

Fig.~2 shows as an example the channel maps of a GMC at 5\,kpc
radius. In the survey it appears as a single entity, with no spatial
or kinematical peculiarities. The high-resolution PdBI map shows
significant substructure at scales of tens of pc and leads to a
different derived virial mass of the complex. Such comparative studies
will help to improve mass determinations.

\section{General Results}

On the basis of H{\sc i} observations (Brinks \& Shane 1984) several
kinematical studies have been undertaken to derive the detailed
properties of the gaseous disk and spiral arms (e.g.\ Braun
1991). This is a rather difficult task because of the thickness of the
H{\sc i} disk, its high inclination and the low contrast of the
``atomic arms''. The survey map of the molecular gas is much better
suited to derive the spiral arm structure and to investigate their
properties. The small width of the ridges indicates a rather short
life time of the molecular gas, but a detailed analysis of the
kinematics has still to be done.

There has been a claim of large streaming motions, deduced from early
CO observations at low angular resolution (Ryden \& Stark 1986). Our
survey data clearly show that the magnitude of non-circular motions in
the molecular gas is only of the order of 10\,km\,s$^{-1}$ which is
also the typical line width. A comparison with the atomic gas shows
that the CO is situated in the middle of the ``atomic arms'', where
the velocity gradient vanishes (Nieten et al., in prep.). The total
range of the widths of individual spectral components is about
4\,\ldots\,15\,km\,s$^{-1}$. However, at many places we
found {\em strong small-scale} disturbances.

Typical examples are double- or multiple-component spectra, broad
lines and short-range spatial variations. From the observational
geometry we are led to conclude that we are not observing chance
line-of-sight coincidences of spatially separated clouds, but true
local effects. Separation of velocity components reach
50\,km\,s$^{-1}$, much more than the intrinsic line width. The sizes
of such regions are of the order of a few primary beams of the PdBI,
hence ideally suited for a detailed observation.

A statistical analysis of the GMCs in the survey as well as a more
detailed investigation of the properties of the clumps in the PdBI
data is under way. Already the few cloud complexes investigated thus
far hint however at important implications for the properties of
molecular gas agglomerations and their analysis: \\[1ex]
$\ast$ CO spectra with multiple components are usually confined
to small regions.\\
$\ast$ The observational geometry indicates a true local phenomenon.\\ 
$\ast$ The substructure of such GMCs consists of well separated
filaments or clumps. \\ 
$\ast$ The velocity separation of the components may be significantly
greater than  \hspace*{1em}the streaming motions or the line width. \\ 
$\ast$ Even seemingly quiescent clouds may consist of individual
clumps which may  \hspace*{1em}afflict the derivation of mass or
temperature. \\[1ex] 
This is the first time that such a wide range of scales of the
molecular gas emission can be investigated and searched for clues
about its state, evolution into stars and mutual interactions with the
other constituents of the ISM.

\acknowledgments 
 
The IRAM technical staff provided excellent receivers, hard- and
software for the new observing modes used. In particular, it is a
pleasure to thank the PdBI staff and astronomers for the superb
observations.

\end{document}